\documentclass[conference]{IEEEtran}
\IEEEoverridecommandlockouts
\usepackage{cite}
\usepackage{amsmath,amssymb,amsfonts}
\usepackage{algorithmic}
\usepackage{graphicx}
\usepackage{textcomp}
\usepackage{xcolor}
\def\BibTeX{{\rm B\kern-.05em{\sc i\kern-.025em b}\kern-.08em
    T\kern-.1667em\lower.7ex\hbox{E}\kern-.125emX}}
\begin{document}

\title{Is Machine Learning Able to Detect and Classify Failure in Piezoresistive Bone Cement Based on Electrical Signals?\\
}

\author{\IEEEauthorblockN{1\textsuperscript{st} Hamid Ghaednia}
\IEEEauthorblockA{\textit{Department of Orthopaedic Surgery} \\
\textit{Massachusetts General Hospital (HMS)}\\
Boston, MA, USA \\
hghaednia@mgh.harvard.edu}
\and
\IEEEauthorblockN{2\textsuperscript{nd} Crystal E. Owens}
\IEEEauthorblockA{\textit{Department of Mechanical Engineering} \\
\textit{Massachusetts Institute of Technology}\\
Cambridge, MA, USA \\
crystalo@mit.edu}
\and
\IEEEauthorblockN{3\textsuperscript{rd} Lily E. Keiderling}
\IEEEauthorblockA{\textit{Department of Orthopaedic Surgery} \\
\textit{Massachusetts General Hospital (HMS)}\\
Boston, MA, USA \\
lkeiderling@mgh.harvard.edu}
\and
\IEEEauthorblockN{4\textsuperscript{th} Kartik M. Varadarajan}
\IEEEauthorblockA{\textit{Department of Orthopaedic Surgery} \\
\textit{Massachusetts General Hospital (HMS)}\\
Boston, MA, USA \\
kmangudivaradarajan@mgh.harvard.edu}
\and
\IEEEauthorblockN{5\textsuperscript{th} A. John Hart}
\IEEEauthorblockA{\textit{Department of Mechanical Engineering} \\
\textit{Massachusetts Institute of Technology}\\
Cambridge, MA, USA \\
ajhart@mit.edu}
\and
\IEEEauthorblockN{6\textsuperscript{th} Joseph H. Schwab}
\IEEEauthorblockA{\textit{Department of Orthopaedic Surgery} \\
\textit{Massachusetts General Hospital (HMS)}\\
Boston, MA, USA \\
jhschwab@mgh.harvard.edu}
\and
\IEEEauthorblockN{7\textsuperscript{th} Tyler N. Tallman}
\IEEEauthorblockA{\textit{School of Aeronautics and Astronautics} \\
\textit{Purdue University}\\
West Lafayette, IN, USA \\
ttallman@purdue.edu}
}

\maketitle

\begin{abstract}
At an estimated cost of \$8 billion annually in the United States, revision surgeries to total joint replacements represent a substantial financial burden to the health care system. 
Fixation failures, such as implant loosening, wear, and mechanical instability of the poly(methyl methacrylate) (PMMA) cement, which bonds the implant to the bone, are the main causes of long-term implant failure. 
Early and accurate diagnosis of cement failure is critical for developing novel therapeutic strategies and reducing the high risk of a misjudged revision. 
Unfortunately, prevailing imaging modalities, notably plain radiographs, struggle to detect the precursors of implant failure and are often interpreted incorrectly. 
Our prior work has shown that the modification of PMMA bone cement with low concentrations of conductive fillers makes it piezoresistive and therefore self-sensing such that when combined with a conductivity imaging modality, such as electrical impedance tomography (EIT), it is possible to monitor load transfer across the PMMA using cost-effective, physiologically benign, and real-time electrical measurements.
Herein, we expand upon these results by integrating machine learning techniques with EIT.
We survey different machine learning algorithms for application to this problem, including neural networks on voltage readings of an EIT phantom for tracking position of a sample, specifying defect location, and classifying defect types. 
Additionally we explore utilization of principal component analysis for each of these problems.
Our results show advantage of neural network with more than 91.9~\%, 95.5~\%, and 98~\% accuracy in interpreting EIT signals for location tracking, specifying defect location, and defect classification respectively. 
Principal component analysis, however, is shown to be an effective addition to machine learning for classifying defect types only.
These preliminary results show that the combination of smart materials, EIT, and machine learning may be a powerful tool for diagnosing the origin and evolution of failure in joint replacement. 
\end{abstract}


\section{\label{sec:Intro}Introduction}
Each year over 2.3 million patients benefit from joint replacement procedures of the knee, hip, shoulder, ankle, and other extremities worldwide \cite{2013Extre,2013HipKnee,kurtz2011international,OrthoWorls}. 
Currently around 6-12\% of all joint replacement procedures involve revisions having a total cost of ~\$8 billion per year in the US \cite{bhandari2012clinical,kurtz2007future}.
Both the replacement and revision operations are estimated to significantly increase by 175\% and 137\%, respectively, by 2030 \cite{pitta2018failure,kurtz2014impact,kenney2019systematic}. 
In particular, patients younger than 55 years of age face an elevated risk of revision due to the greater demands placed on their joints as well as the steady increase in risk of implant fixation failure with in vivo duration \cite{AustralianOrthopaedic,NationalJointRegEng}.
In the long-term, different failure mechanisms of fixation between implant and bone, such as aseptic loosening and instability, account for more than 60\% of the revision cases.
The fixation between implant and bone is provided either by biological fixation (cementless implants) or by poly(methyl methacrylate) (PMMA) usually called bone cement.
Cemented implants are the standard of care for knee, elbow, and shoulder arthroplasty and are reclaiming their popularity compared to cementless fixation for total hip replacement  \cite{no201714th,nam2019cemented,lawrie2019cost}. 

Despite the significant efforts in the field, much uncertainty remains regarding the mechanisms by which fixation failures evolve \cite{jacobs2001osteolysis,kenney2019systematic,pitta2018failure,postler2018analysis}.
The key constraint in understanding these challenges is the lack of an early diagnosis and predictive method.
The importance of this clinical issue is perhaps best illustrated by an example. 
In early 2017, Bonutti et al. \cite{bonutti2017unusually} reported high rates of early tibial component loosening in patients with a new implant design (15 knees, < 2 years post-surgery). 
Radiographic evaluation demonstrated loosening in only 2 out of the 15 knees. 
However, intra-operatively all knees were found to have grossly loose implants requiring revision. 
The challenge is that failure initiates well before indicators become visible.

\begin{figure}[!]
\centering
\includegraphics[width=0.45\textwidth]{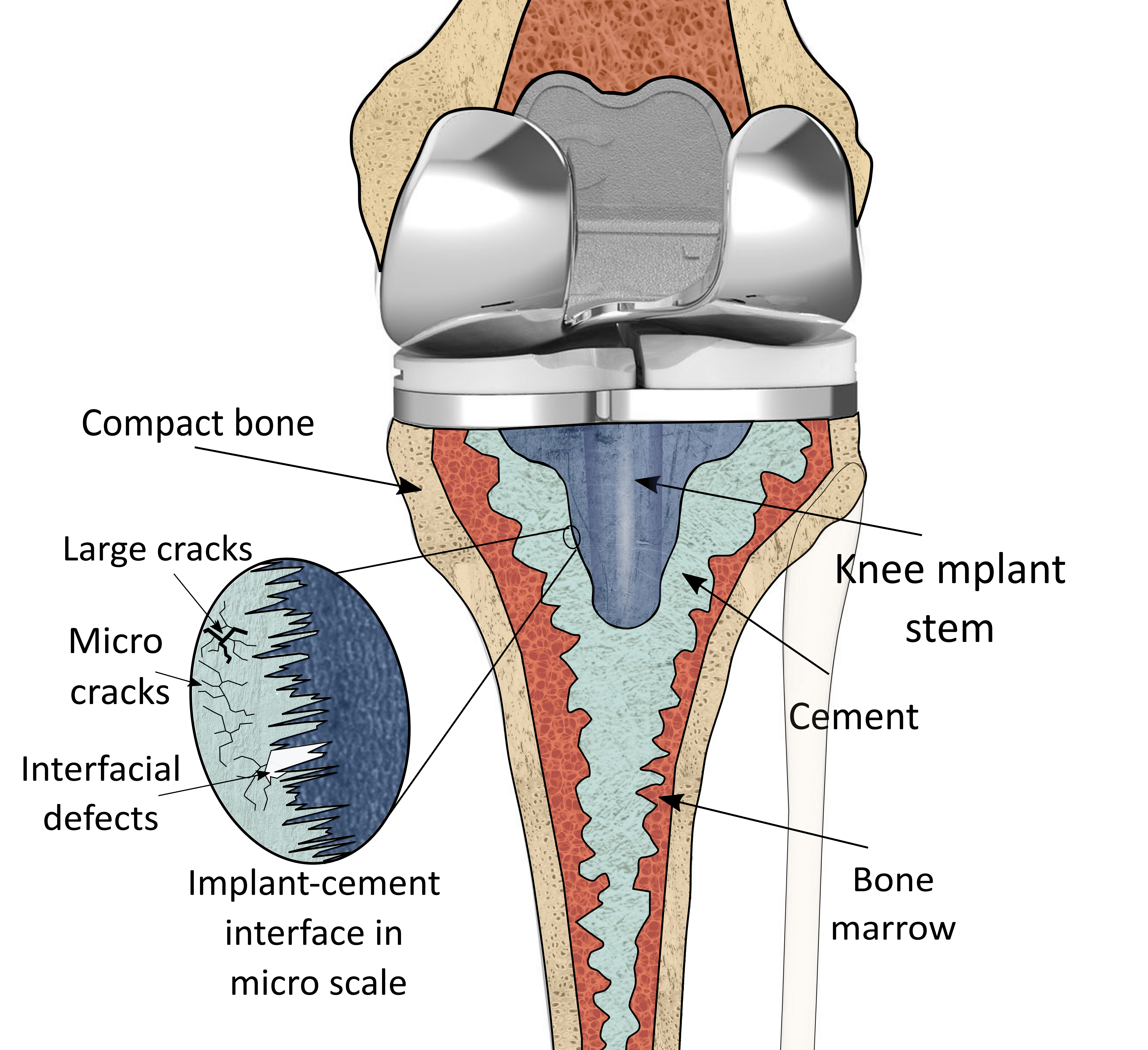}
\caption{\label{fig:Fig1} The schematic of the bone-implant-cement interface in a knee replacement. The mechanical failures start with initiation of micro cracks within the bone cement. The propagation of these micro-crack can cause interfacial defects or fracture.}
\end{figure}

Our hypothesis is that, similar to interface failure between any two materials, the mechanical failure for cemented joint replacements initiates with the propagation of microcracks within the bone cement caused by contact/interfacial stresses \cite{ghaednia2017review,ghaednia2016comprehensive}.
The contact stresses are maximum beneath the interface (within the cement) rather than at the interface. This causes the initiation of micro-cracks within the bone cement volume as illustrated in Fig.~\ref{fig:Fig1}\cite{ghaednia2017review}.
Even though at this stage the defects are not detectable by conventional diagnosis devices, the stress distribution within the bone cement is altered by the micro-cracks.
The accumulation and propagation of these micro cracks continues and can yield two failure mechanisms: i) when cracks reach the interface and cause loss of fixation (aseptic loosening) or ii) as the micro-cracks merge they create larger cracks and hence breakage of the fixation.
At this stage the defects are visible however the failure is already in advance stages and will require revision.  
It is for these reasons that traditional imaging modalities such as plain radiographs, CT scans, arthrography, and nuclear imaging are largely limited to diagnosis of mechanical failures in advanced stages \cite{rea2007radiolucency,torrens2009assessment}. This also provides powerful motivation for the development of new assessment modalities of this important clinical problem.

Motivated by this challenge, our previous work proposed a new method for in vivo monitoring failure in cemented orthopaedic implants \cite{ghaednia2020interfacial2SMS,ghaednia2019noninvasive,Ghaednia2020}. 
The conceptual framework of our approach is to modify the PMMA with conductive nano-to-microscale fillers such that it becomes self-sensing via the piezoresistive effect (i.e. having deformation and damage-dependent electrical conductivity). Benign electrical measurements collected at the skin of the patient can then be used to monitor the stress/strain state and overall health of the cement interface.
Specifically in our prior work \cite{ghaednia2020interfacial2SMS}, PMMA bone cement was modified with low volume fractions of chopped micro carbon fiber (CF) to impart piezoresistive-based self-sensing. 
Because the conductivity of this material system is percolation-derived, interfacial stress and deformations, which alter the connectivity of the nanofiller network, manifest as a conductivity change. 
Electrical impedance tomography (EIT) was then used to detect and monitor load-induced deformation and fracture of CF/PMMA in a phantom tank. 
Our results showed that EIT is indeed able to detect compressing force on a prosthetic surrogate, distinguish between increasing load magnitudes, detect failure of implant fixation, and even differentiate between cement cracking and cement de-bonding without direct contact with the surrogate. 
In this paper we expand upon our previous work to investigate the potential of failure classification via machine learning.

Monitoring conductivity changes due to mechanical strains and defects in piezoresistive materials via EIT has precedent in structural venues such as structural health monitoring (SHM) and nondestructive evaluation (NDE) \cite{tallman2016damage,tallman2015damage,smyl2018detection}.
In these applications, EIT has proven to be capable of identifying and localizing the location of impact damages as well as elastic and plastic deformations. Due to its low spatial resolution compared to radiographic imaging, however, EIT has yet to see wide-spread embrace in health care. 
Nonetheless, the cost effectiveness, benign nature, and ultra-high temporal resolution of EIT make it an appealing technology. To this end, there have been several recent efforts for using machine learning, especially neural networks, in reconstruction of conductivity maps via EIT measurements \cite{wang2004rbf,rymarczyk2018non,klosowski2017using,khan2019review,cao2013direct,hamilton2018deep,lee2013functionalizing,martin2015electrical,martin2017post,ren2019statistical,wang2018unsupervised,wei2019dominant}.
In this work, instead of reconstructing conductivity maps with ML, we solve either a classification or a regression problem thus reducing the number of outputs from on the order of hundreds (i.e. the number of elements in the finite element mesh onto which EIT endeavors to reproduce the conductivity distribution) to only a few -- namely, the defect characteristics. 
Specifically, we investigate viability of machine learning for three different cases: tracking location of an object in a domain, specifying orientation of a defect in a piezoresistive bone cement, and classifying defect types or health conditions of a piezoresistive bone cement.

\section{\label{sec:Mat}Materials and Methods}

\subsection{\label{subsec:Cement} Piezoresistive Cement}
\begin{figure}[!]
\centering
\includegraphics[width=0.45\textwidth]{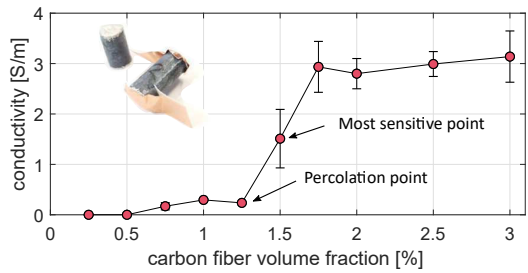}
\caption{\label{fig:Fig2} Conductivity vs. CF-PMMA vol.\% of cylindrical samples with different CF-PMMA vol.\% to determine the percolation behavior and most sensitive point to loading increase and decrease}
\end{figure}

As described in our previous work \cite{ghaednia2020interfacial2SMS}, piezoresistive bone cement (pBC) was formulated by mixing different amounts of chopped CF with aspect ratio of 430 (3 mm length and $7~\mu$m diameter; TEJIN Carbon America Inc., Rockwood, TN, USA) in dental PMMA cement (Bosworth Fastray). 
CFs were first tip sonicated in ethanol at 30~W and 20~kHz for an hour to disperse the fibers. 
The solution was then left to dry slowly to a powder on a heater plate at 65$^\circ$C for 3 days. 
The CFs were then added to the dry PMMA, which included benzoyl peroxide as a radical initiator, and gently mixed using mortar and pestle.
Methyl methacrylate (MMA) monomer was added to the dry mixture (800 ml MMA/gr of PMMA) which crosslinked with the powder to form a polymerized solid material. 
Cylindrical CF/PMMA specimens were molded with 10 different volume fractions of CF-to-PMMA ranging from 0.1\% to 3.0\%. 
In order to measure conductivity, the bottom and top faces of the cylindrical specimens were polished, cleaned with ethanol, painted with silver conductive paint, and covered with copper tape to reduce contact resistance, Fig.~\ref{fig:Fig2}. 
Our study showed that percolation and maximum gradient occur near 1.25~vol.\% CF and 1.5~vol.\% CF respectively, Fig.~\ref{fig:Fig2}. The same recipe with 1.5~vol.\% CF was used to build the samples for this work.

\begin{figure}[!]
	\begin{center}
		\includegraphics[width=0.35\textwidth]{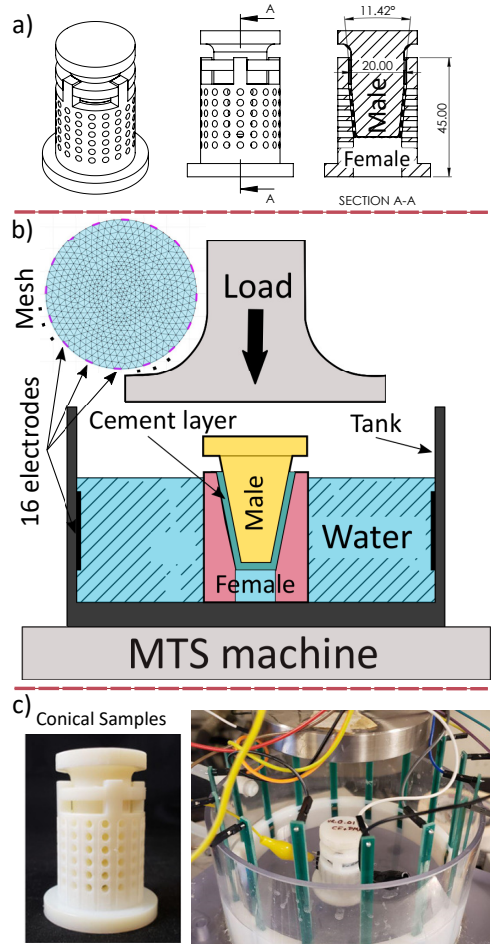}
		\caption{\label{fig:Fig3} a) Drawings of the conical samples (dimensions in mm), b) Schematic of the EIT setup and the mesh used for calculating the conductivity map, c) Left: 3D printed conical samples, right: EIT phantom under MTS machine.}
	\end{center}
\end{figure}

\subsection{\label{subsec:EIT}Electrical Impedance Tomography Phantom}

An EIT phantom consisted of a cylindrical tank filled with deionized water and 16 electrodes spaced evenly around the exterior of the domain, Fig.~\ref{fig:Fig3}(b,c) \cite{ghaednia2019noninvasive,ghaednia2020interfacial2SMS,Tallman2019}. 
Test specimens consisting of a conical female component, which emulated the cavity within long bones (e.g. a femur), and a conical male component coated with silver paint, which simulated the surface of metal prosthesis (e.g. a femoral stem), were 3D printed, Fig.~\ref{fig:Fig3}(a,c). 
A layer of pBC was sandwiched between the male and female components, Fig.~\ref{fig:Fig3}(b). 
The male component was then subjected to compressive loads using a MTS machine Fig.~\ref{fig:Fig3}(b,c) while EIT measurements were collected from a 16-electrode array encircling the phantom.
Loads were increased incrementally from 45 N to 4000 N.
EIT measurements were then post-processed via an in-house EIT routine developed by Tallman et al. to reconstruct the conductivity maps \cite{tallman2017effect}.

\subsection{\label{subsec:EIT}Machine Learning}

Herein, we utilized existing machine learning (ML) algorithms onto two different data set types.
In the first approach, we applied machine learning directly to the raw voltage signals, which included all available data, but risked overfitting due to a large number of features.
In the second approach, we first applied principal component analysis (PCA) on the voltage signals to reduce the number of features, and subsequently applied machine learning on only the first few principal components that captured more than 95\% of the data variability.
Matlab machine learning and neural network tool boxes were used to train different machine learning algorithms on the experimental data and create confusion matrices.
As part of this study, we compared machine learning methods such as support vector machine (SVM), K-Nearest Neighbor (KNN), trees, and ensemble methods, to a two-layer feed-forward neural network for both classification and regression problems based on EIT measurements, in order to identify the most efficacious methods for these problems and for the low level of data available relative to what machine learning methods typically require for optimal performance.

\section{\label{sec:ExpRes}Results and Discussion}
Three sets of experiments were designed and performed: 1) tracking the location of a sample within a tank; 2) specifying the orientation of a vertical crack in a conical sample; and 3) classifying health conditions (i.e. healthy, vertically cracked, horizontally cracked, or loose/disbonded). These experiments were designed to investigate the viability of machine learning for interpreting EIT measurements.

\subsection{\label{subsec:Loc}Location Tracking}

\begin{figure}[h]
	\centering
	\includegraphics[width=0.45\textwidth]{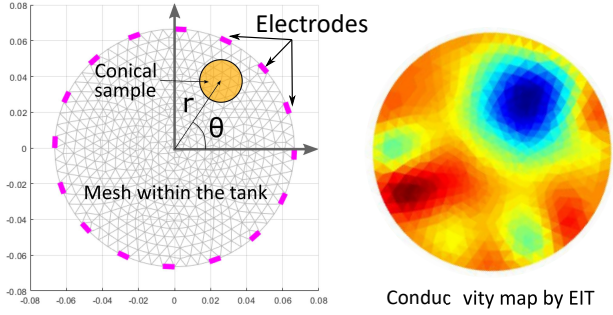}
	\caption{\label{fig:Fig4} EIT measurements were performed on the conical samples in 112 different locations within the tank.}
\end{figure}

In the first set of experiments, we placed a conical specimen (Fig.~\ref{fig:Fig3}(a)) in different locations within the EIT phantom tank. 
The radial coordinates, $r$ and $\theta$, of the specimen center were tracked, Fig.~\ref{fig:Fig4}.
Specimens were placed at $r =$ 0~cm, 2~cm and 4~cm and at $\theta =  0^\circ:30^\circ:330^\circ$.
At each location, EIT measurements were performed on four different conical specimens and with 100 recordings at each measurement for averaging.
We used the data from three specimens for the training whereas one specimen was used solely for testing.
For this set of tests, analytical EIT solutions for reconstructing the phantom conductivity map were able to track the location of the specimen, Fig.~\ref{fig:Fig4}.

\begin{figure}[h]
	\includegraphics[width=0.5\textwidth]{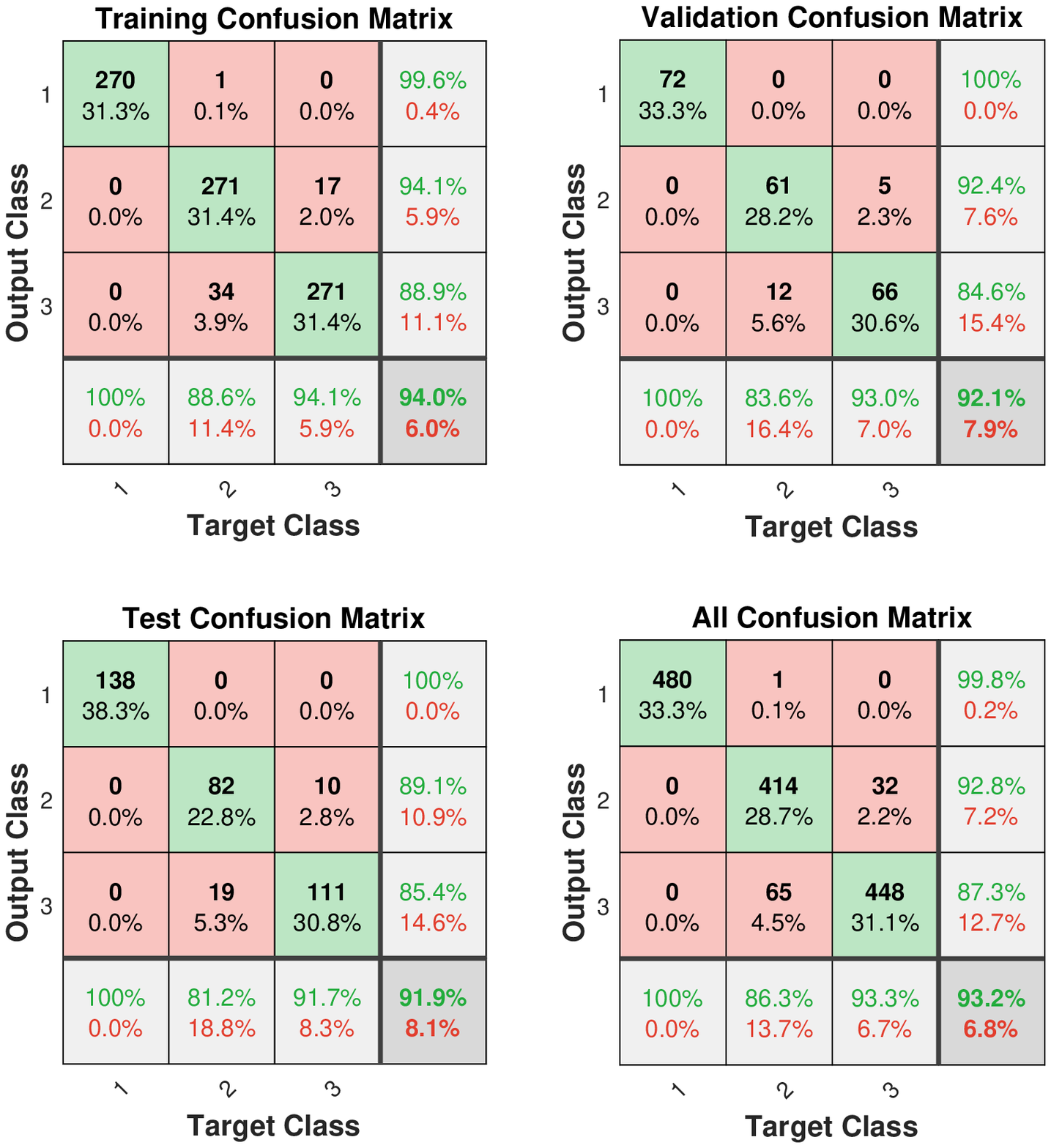}
	\caption{\label{fig:Fig5} Confusion matrices for classifying the radial distance of conical samples, using MATLAB's neural network pattern recognition toolbox.}
\end{figure}

We compared different machine learning algorithms to predict radial coordinates components $r$ and $\theta$, Fig.~\ref{fig:Fig4}. 
Here, 60\% of the data was used for training, 15\% for validation and 25\% for testing. 
Using an SVM algorithm directly on the EIT measurements resulted in 100\% training and validation and 92\% training accuracy for classifying the radial position of the sample.
For the angular component, $\theta$ in Fig.~\ref{fig:Fig4}, SVM resulted in 100\% training accuracy and validation and 83\% testing.
However, due to the small number of tests compared to the large number of features, there is a high risk of overfitting.

\begin{figure}[h]
	\includegraphics[width=0.5\textwidth]{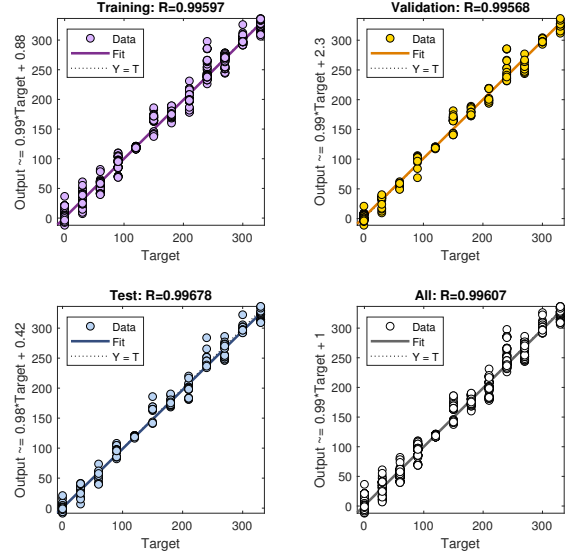}
	\caption{\label{fig:Fig6} The regression fit of the angular component, $\theta$, using MATLAB's neural network fit toolbox.}
\end{figure}

Using the data pretreatment, the first four principal components described 95.5\% of the data variability. 
Even though the training and validation of the regular machine learning algorithms resulted in 93\% accuracy for the SVM algorithm, the testing of the trained model was very low at 39\%. 
Other linear and nonlinear algorithms, such as decision trees and KNN, also worked well for training but not for testing. These results were similar for learning the angular component of position, $\theta$.

\begin{figure*}[!]
	\includegraphics[width=.95\linewidth]{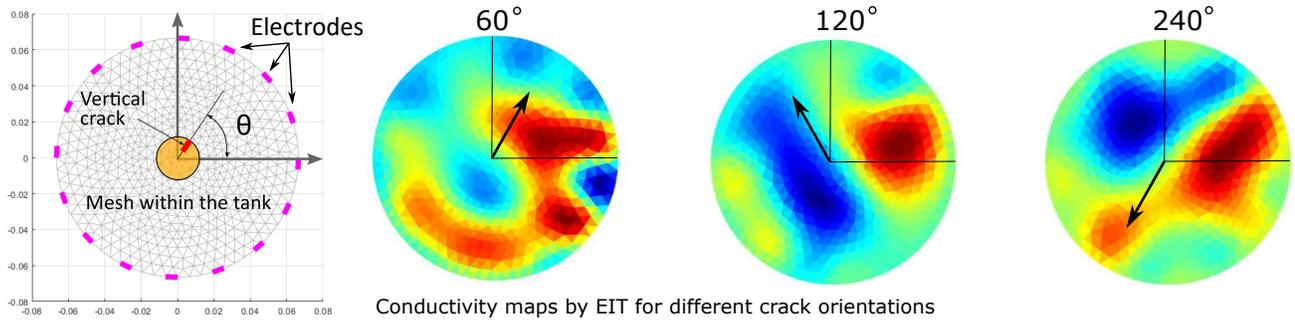}
	\caption{\label{fig:Fig7} A vertically cracked conical sample was placed at the middle of the tank. The sample was rotated to face 12 different angles.}
\end{figure*}

Next, we used two layer feed-forward neural networks with five hidden neurons, and directly used pretreated data. 
For this, the MATLAB's Neural Network Pattern Recognition toolbox was used.
The first four PCAs were used to train the neural networks.
We again used 60\% of the data for training, 15 percent for validation of the training and 25\% for testing. 
The confusion matrices for training, validation, and testing of the model for classifying the radial distance are shown in Fig.~\ref{fig:Fig5}.
Overall, the neural network resulted in 94.0\% accuracy in training, 92.1\% in validation, and 91.9\% in testing.
For the angular component, $\theta$, we used MATLAB's neural network fit toolbox.
Similar to the classification of the radial component, this resulted in acceptable results in training, validation and testing, with an average of $10^\circ$ error, see Fig.~\ref{fig:Fig6}.

This analysis shows that machine learning algorithms, especially neural networks, are able to interpret the EIT measurements for tracking location of an object inside a domain; however, we did not see an improvment in accuracy of machine learning algorithms when principal component analysis was used for tracking the location of samples.

\subsection{\label{subsec:CrackOri}Crack Orientation}

We designed a second set of experiments to investigate if machine learning is better able to interpret boundary voltage data than analytical optimization-based solutions in order to locate defects on a sample.
In this experiment set, a 1 mm width vertical cut was made by a bandsaw on four conical specimen that were placed at the center of the phantom tank.
The specimen were rotated from $0^\circ$ to $360^\circ$ by $30^\circ$ steps with a fixed location at the center of the phantom tank.
At each angle, EIT measurements were performed with 100 readings of signals at each measurement for averaging.
A reconstruction the conductivity map using analytical EIT solution failed to distinguish between crack orientations as shown by the representative set of EIT-generated conductivity maps in Fig.~\ref{fig:Fig7}.
The EIT maps in Fig.~\ref{fig:Fig7} shows the conductivity difference between crack facing an angle $\theta$ and baseline of $\theta = 0$. 
Similar to location tracking, we used the data from three specimens for the training whereas one specimen was used solely for testing.
Therefore, 60\% of the data was used for training, 15\% for validation, and 25\% for testing the model.
The linear SVM algorithm applied directly to the EIT measurements resulted in 100\% training and validation and 83\% testing accuracy.
However, there is again a high risk of over fitting due to the large number of features compared to experiments.

\begin{figure}[!h]
	\centering
	\includegraphics[width=0.45\textwidth]{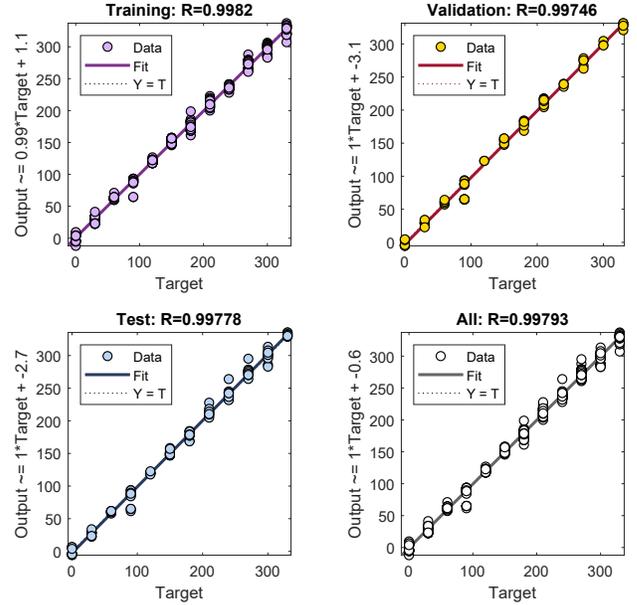}
	\caption{\label{fig:Fig8} Neural network regression fit for calculating crack orientation using the Levenberg-Marquardt algorithm.}
\end{figure}

We also investigated using principal components of the voltage signals instead of the raw voltages.
The first four principal components showed 95.5\% of the data variably and were used for developing the following machine learning algorithms.
The Gaussian Process Regression available in the regression learner toolbox of MATLAB resulted in $5.6^\circ$ root mean square error during the training and $16.73^\circ$ during the testing.
Next we used neural network regression fit available in MATLAB neural network toolbox. 
The first four principal components were used as the model features. Again, 60\% of data was used for training, 15\% for validation, and 25\% testing the model.
This resulted in root mean square errors of $6.1^\circ$ and $7.0^\circ$ for training and testing, respectively.
Even with a small numbers of experiments, this suggests that machine learning is able to track the defect orientation thereby demonstrating the burgeoning potential of using machine learning to better interpret EIT measurements.

\subsection{\label{subsec:Defclass}Failure Classification}

In the third set of experiments, we considered four cement health conditions: healthy, vertically cracked, horizontally cracked, and loose, Fig.~\ref{fig:Fig9}(a). Three specimens were tested for each condition.
The cracks were implemented as a bandsaw cut in both the surrogate and the cement layer.
For looseness, we coated the male components of the conical specimens with Vaseline before curing the cement to allow conformal molding but prevent adhesion. Looseness was confirmed by manually pulling out the male component after the cement cured.
All of the specimens were prepared with 1.5 vol.\% CF pBC \cite{ghaednia2020interfacial2SMS}.
Each specimen was then subjected to compression loading from 300 to 2200 N as EIT measurements were simultaneously collected.
The conductivity change measured by EIT showed consistent increase in the absolute conductivity change for all of the specimens, Fig.~\ref{fig:Fig9}(b).
Healthy specimens show the largest change followed by the loose specimens, vertically cracked specimens, and, lastly, horizontally cracked specimens.

\begin{figure}
\includegraphics[width=0.45\textwidth]{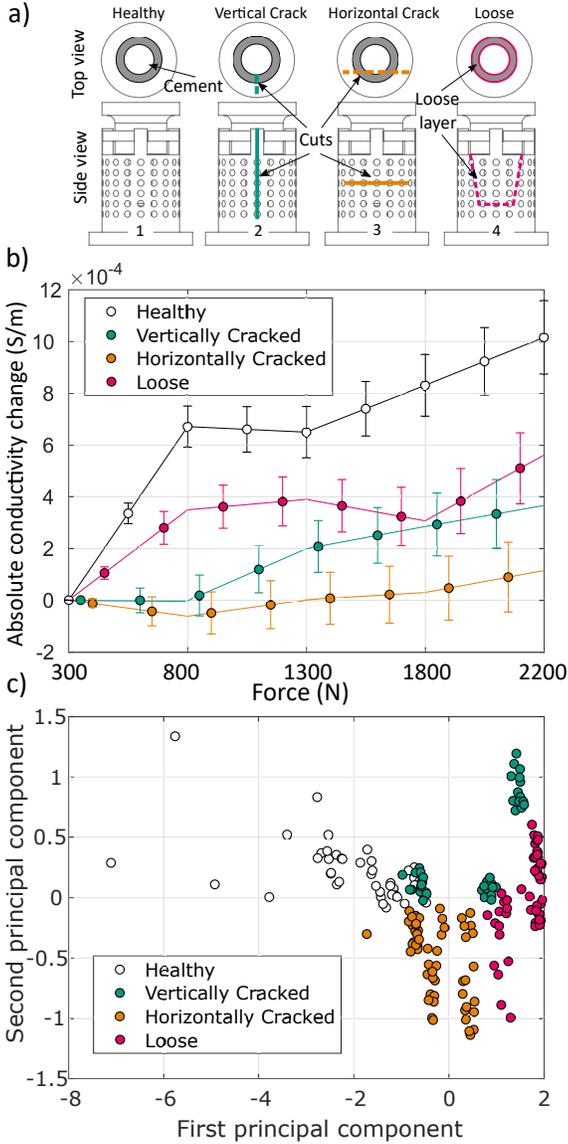}
\caption{\label{fig:Fig9} a) Defect types. b) conductivity change as a function of compression loading measured by EIT. c) PCA of the signals showing four distinct clusters.}
\end{figure}

Because of the low number of tests in this set of experiments relative to the number of health conditions, we immediately used EIT data pretreated with PCA.
Contrary to the prior two cases, location tracking and crack orientation, the first four principal components found here described only 88\% of the data variability. 
To cover 95\% variability, seven principal components had to be considered; however, for consistency and to avoid potential overfitting we decided to use the first four principal components for this analysis.
Fig.~\ref{fig:Fig9}(c) shows the first two principal components for all specimens and all loading conditions. 
The results show four divided clusters for each health condition with the exception of a few similarities between vertically cracked and healthy samples.
We then trained different machine learning algorithms with five-fold cross validation using only the first four principal components of the EIT measurement on 75\% of the data, and reached 99.3\% accuracy in training and 98\% testing accuracy (25\% of the data) for classifying failure types using the KNN algorithm.
We also used MATLAB's two layer feed-forward neural network with three hidden neurons. 
The neural network resulted in 100\% accuracy in training and validation and 96.2\% in testing accuracy, which was expected considering how the principal components showed four well-divided clusters for the health conditions.
This result demonstrates the viability of using PCA with multiple machine learning methods to classify defect types in a piezoresistive bone cement, even given the limited number of experiments.

\begin{figure}
\includegraphics[width=0.5\textwidth]{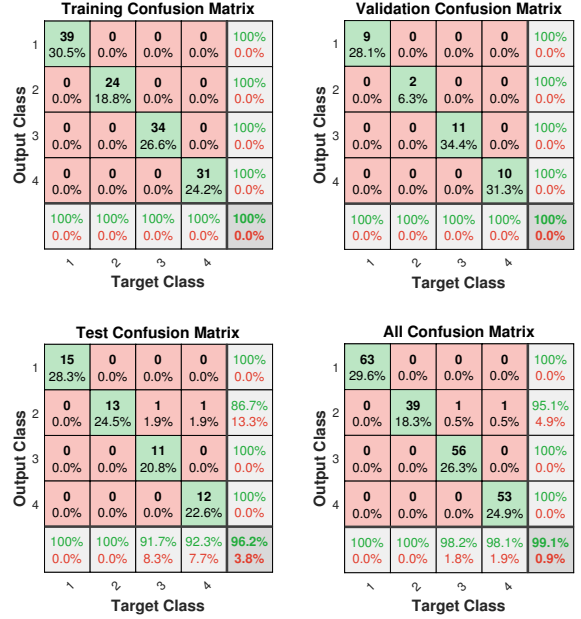}
\caption{\label{fig:Fig9} Confusion matrices for training a two layer feed-forward neural net for classifying the health condition of piezoresistive samples.}
\end{figure}


\section{\label{sec:conc}Conclusion}

In this study we investigated the viability of data analysis techniques, including principal component analysis and machine learning, in interpreting EIT measurements for application to monitoring of cemented joint replacement health and condition, using piezoresistive bone cement.
We designed three sets of experiments to analyze different aspects of this problem.
The first set of experiments was designed to verify whether machine learning was able to localize conductivity-changing effects, such as objects, which is a task that analytical EIT solutions can perform well.
To test this, we tracked the position of a joint replacement surrogate specimen in a phantom tank by applying machine learning directly to the EIT measurements and onto measurements pretreated by principal component analysis.
A simple neural network was able to successfully track the radial coordinates of the position for both data types.
The second set of experiments evaluated machine learning methods to classify subtleties embedded in EIT measurements that may be missed by traditional EIT formulations due to regularization. 
For this set, we placed a vertically cracked specimen at the center of a water tank and rotated it to move the crack.
While EIT was not able to specify the orientation of the crack, machine learning methods including a two layer neural network including 5 neurons, SVM, KNN and ensemble methods were able to determine crack orientation with reasonable accuracy.
The third set of experiments tested the viability of combining machine learning and EIT for diagnosis of defects. 
This experiment was designed to test the potential of this method for monitoring joint replacement defects following replacement using cemented implants. 
We investigated this by testing four health conditions: healthy, vertically cracked, horizontally cracked, and entirely loose samples.
The first two principal components alone showed divided clusters for each health condition, which facilitated training of machine learning algorithms.
Finally, we showed two machine learning methods, KNN and neural networks, were both able to interpret EIT measurements even with a relatively small number of training sets. 

In conclusion, the combination of EIT, machine learning, and self-sensing bone cement appears to have considerable potential for solving the important clinical problem of correctly assessing failure in cemented total joint replacements. It should be noted, however, that this proof-of-concept study and machine learning method survey should be expanded in future work. Specifically, future work should consider more realistic damage cases (i.e. real fatigue cracks as opposed to saw cuts) and application in the presence of native physiological tissue via cadaver testing. Physiological compatibility of the modified PMMA bone cement is also an important consideration.


\section{\label{sec:ackn}acknowledgments}
Financial support was provided by the MGH/MIT Strategic Partnership Grand Challenge Grant (to A.J.H. and K.M.V.), and the Department of Defense through the National Defense Science and Engineering Graduate Program (to C.E.O.).

\bibliographystyle{acm}
\bibliography{EIT_ML_Paper}

\end{document}